\documentclass[aip,cha,nofootinbib,twocolumn,showkeys,showpacs,floatfix,10pt]{revtex4-1}

\usepackage{ifpdf}
 \ifpdf
   	\usepackage[T1]{fontenc}  
   	\usepackage{lmodern}
   	\usepackage[pdftex]{hyperref,graphicx}	% using pdflatex 
    	\hypersetup{colorlinks,% 
    	citecolor=blue,% 
    	filecolor=green,% 
    	linkcolor=blue,% 
    	urlcolor=blue,% 
   	}
 \fi

\usepackage{xcolor}
\usepackage[english]{babel}
\usepackage{cleveref}
\usepackage{booktabs}
\usepackage[utf8x]{inputenc}
\usepackage[normalem]{ulem}
\usepackage{microtype}
\usepackage{subfigure}
\usepackage{bm}% bold math
\usepackage{times}
\usepackage{amsmath, amssymb}

\newcommand{\half}[0]{\frac{1}{2}}
\newcommand{\bigO}[0]{\mathcal{O}}
\newcommand{\SSzero}{\rm {\bf SS}$_0$}
\newcommand{\DSone}{\rm {\bf DS}$_{\textnormal{SN}^-}$}
\newcommand{\DStwo}{\rm {\bf DS}$_{\textnormal{SN}^+}$}
\newcommand{\DSthree}{\rm {\bf DS}$_{\textnormal{U}}^1$}
\newcommand{\DSfour}{\rm {\bf DS}$_{\textnormal{U}}^2$}
\newcommand{\DSLC}{\rm {\bf DS}$_{\textnormal{LC}}$}
\newcommand{\SDLC}{\rm {\bf SD}$_{\textnormal{LC}}$}
\newcommand{\SDone}{\rm {\bf SD}$_{\textnormal{SN}^-}$}

\newcommand{\DD}{\rm {\bf DD}}
\newcommand{\DDeq}{\rm {\bf DD}$_{\textnormal{TC}}$}
\newcommand{\DDlc}{\rm {\bf DD}$_{\textnormal{H}}$}
\newcommand{\DS}{\rm {\bf DS}}
\newcommand{\SD}{\rm {\bf SD}}
\newcommand{\SSpi}{\rm {\bf SS}$_\pi$}
\newcommand{\SSzeroI}{\text{\textbf{SS}}_0}

\newcommand{\SSpiI}{\text{\textbf{SS}}_\pi}
\newcommand{\I}{\rm {\bf I}}
\newcommand{\II}{\text{\textbf{I}}}

\newcommand{\ud}{\textrm{d}}

\newcommand{\Tor}{\mathbb{T}}
\newcommand{\R}{\mathbb{R}}
\newcommand{\Z}{\mathbb{Z}}

\newcommand{\set}[2]{\rset{#1}{#2}}
\newcommand{\sset}[1]{\left\lbrace #1\right\rbrace}

\newcommand{\rset}[2]{\left\lbrace\, #1\,\left|\;#2\right.\right\rbrace}

\newcommand{\abs}[1]{\left|#1\right|}

\DeclareMathOperator{\Fix}{Fix}
\DeclareMathOperator{\Img}{Im}
\DeclareMathOperator{\Rep}{Re}
\newcommand{\Ss}{\mathcal{S}}
\renewcommand{\SS}{\Ss\Ss}

\newcommand{\Sot}{\Sigma_{21}}
\newcommand{\RR}{\mathcal{R}}

\newcommand{\CC}{\mathcal{C}}

\definecolor{colEM}{rgb}{.9,0,0}
\definecolor{colMP}{rgb}{0,.7,0}
\definecolor{colCB}{rgb}{0,0,.7}

\renewcommand{\Re}{\Rep}

\begin{document}

% ----------------------------------------------------------------------
\title{Chimera states in two populations with heterogeneous phase-lag} %Title of paper
% ----------------------------------------------------------------------
\author{Erik A.~Martens}%
\email{erik.martens@ds.mpg.de}
\homepage{http://eam.webhop.net}
\affiliation{Dept. of Biomedical Sciences, University of Copenhagen, Blegdamsvej 3, 2200 Copenhagen, Denmark}
\affiliation{Dept. of Mathematical Sciences, University of Copenhagen, Universitetsparken 5, 2200 Copenhagen, Denmark}

\author{Christian Bick}%
\affiliation{Dept. of Mathematics, University of Exeter, Exeter, United Kingdom}

\author{Mark J.~Panaggio}%
\affiliation{Mathematics Dept., Rose-Hulman Institute of Technology, Terre Haute, IN, USA}
\affiliation{Dept. of Engineering Sciences and Applied Mathematics, Northwestern University, Evanston, IL, USA}
% ----------------------------------------------------------------------
%     Following are some guidelines for processing of LaTeX Article Files on the journal's submission & peer-review system:
%         A LaTeX submission should be a single .tex file; multiple-(.tex)-file submissions are not supported.
%         Uploading .bib files as Additional LaTeX file types is supported.
%         Commands to include figures may be used. Ensure that the figure filename cited in the command matches that of the actual file upload; use only the simple filename, not a complete directory path; only include figures in .eps format (.jpg,.pdf, and .tif while acceptable will not appear inline). If an EPS file contains a thumbnail preview, it may not render in the article PDF generated via the TeX.
%         JPG and PDF figures are acceptable for submission but will only appear inline if you have defined the bounding box area.
%         Footnotes should be treated the same as references. Do not use the "\footnote" command; use the standard "\cite" command for footnotes as well as references.
%         Processing and conversion of LaTeX files generated from Scientific Word (tcilatex) is not supported, and not likely to proceed successfully.
%         Submissions in Plain TeX format or in dvi format are not supported.
%         To redo the TeX processing, e.g., after uploading a new or revised figure file, do a Replace on the .tex Article File. Doing a Rebuild of the Merged PDF file does NOT initiate another TeX processing.
% ----------------------------------------------------------------------
\date{\today}
% ======================================================================
\begin{abstract}
The simplest network of coupled phase-oscillators exhibiting chimera states is given by two populations with disparate intra- and inter-population coupling strengths. We explore the effects of heterogeneous coupling phase-lags between the two populations. Such heterogeneity arises naturally in various settings, for example as an approximation to transmission delays, excitatory-inhibitory interactions, or as amplitude and phase responses of oscillators with electrical or mechanical coupling. 
We find that breaking the phase-lag symmetry results in a variety of states with uniform and non-uniform synchronization, including in-phase and anti-phase synchrony, full incoherence (splay state), chimeras with phase separation of $0$ or $\pi$ between populations, and states where both populations remain desynchronized. These desynchronized states exhibit stable, oscillatory, and even chaotic dynamics. 
Moreover, we identify the bifurcations through which chimeras emerge.
Stable chimera states and desynchronized solutions, which do not arise for homogeneous phase-lag parameters, emerge as a result of competition between synchronized in-phase, anti-phase equilibria, and fully incoherent states when the phase-lags are near $\pm\frac{\pi}{2}$ (cosine coupling).
These findings elucidate previous experimental results involving a network of mechanical oscillators 
and provide further insight into the breakdown of synchrony in biological systems.
\end{abstract}
% ======================================================================
\pacs{05.45.-a, 05.45.Xt, 05.65.+b}% PACS, the Physics and Astronomy Classification Scheme.

\keywords{chimera states, phase-lag, hierarchical network, neural networks}

\maketitle %\maketitle must follow title, authors, abstract and \pacs

\begin{quotation}
The synchronization of oscillators is a ubiquitous phenomenon that manifests itself in a wide range of biological and technological settings, including the beating of the heart~\cite{Michaels1987}, flashing fireflies~\cite{Buck1968}, pedestrians on a bridge locking their gait~\cite{Strogatz2005}, circadian clocks in the brain~\cite{Liu1997}, superconducting Josephson junctions~\cite{Wiesenfeld1998}, chemical oscillations~\cite{Kiss2002,Taylor2009}, metabolic oscillations in yeast cells~\cite{Ghosh1971,Dano1999}, and life cycles of phytoplankton~\cite{Massie2010}. Recent studies have reported the emergence of solutions where oscillators break into localized synchronized and desynchronized populations, commonly known as chimera states~\cite{Kuramoto2002,Panaggio2015b}. These solutions have been studied in the Kuramoto-Sakaguchi model with homogeneous coupling phase-lag~\cite{Abrams2008,Montbrio2004,MartensPanaggioAbrams2015,Panaggio2016}. Significant progress has been made understanding how chimera states emerge with respect to different topologies~\cite{Abrams2004,Panaggio2013,Panaggio2015a,Maistrenko2015}, their robustness towards heterogeneity~\cite{Laing2009, Laing2012a}, how they manifest in real-world experiments such as (electro-) chemical and mechanical oscillator systems~\cite{Tinsley2012,MartensThutupalli2013,Wickramasinghe2013} and laser systems~\cite{Hagerstrom2012}, and recently in explaining their basins of attraction~\cite{MartensPanaggioAbrams2015} and controllability~\cite{BickMartens2014,MartensPanaggioAbrams2015}. 
Here we generalize one of the simplest systems in which chimera states are known to occur, two populations of identical phase-oscillators with heterogeneous intra- and inter-population coupling, to account for effects of breaking the symmetry in the phase-lag parameters.  Using symmetry considerations, numerical methods and perturbative approaches, we explore and explain the emergence of dynamics which only occur for heterogeneous phase lags, including new types of chimera states and desynchronized attractors with stable, periodic, or chaotic motion. We find that equilibria with non-uniform synchrony such as chimeras are stable near four points in parameter space where time-reversing symmetries exist and where fully synchronized in-phase and anti-phase states and fully incoherent states exchange stability. These findings corroborate the notion that chimera states emerge as a competition between different types of uniform synchronization~\cite{MartensThutupalli2013}.
\end{quotation}

% ==============================================
\section{Introduction}
% ==============================================

% General Introduction:
Over a decade ago, the observation of solutions characterized by localized synchrony and incoherence~\cite{Kuramoto2002}, which subsequently became known as chimera states~\cite{Abrams2004}, sparked an enormous amount of interest in coupled oscillatory systems. For identical oscillators, such dynamics exhibit symmetry breaking: the solution has less symmetry than the system itself~\cite{Bick2015d}. At the same time, chimera states are robust against heterogeneities, including additive noise, non-identical oscillator frequencies~\cite{Laing2009}, various coupling topologies~\cite{Shima2004,Martens2010swc,Panaggio2013,Panaggio2015a,Maistrenko2015}, and non-complete network topologies~\cite{Laing2012a}. 
They have since been observed in real-world systems such as experimental systems ranging from metronomes~\cite{MartensThutupalli2013} to (electro-)chemical oscillators and lasing systems\cite{Wickramasinghe2013,Tinsley2012, Hagerstrom2012, Schonleber2014}. Moreover, by applying control, they may be relevant for functional applications in neurobiology~\cite{BickMartens2014,Omelchenko2016,MartensPanaggioAbrams2015}. For a detailed review on chimera states, see Ref.~\cite{Panaggio2015b}.

One of the simplest models in which chimera states arise consists of two interacting populations composed of~$N$ Kuramoto--Sakaguchi phase oscillators, where the phase $\theta_k^\sigma\in\Tor = \R/2\pi\Z$ of the $k$th oscillator in population $\sigma=1,2$ evolves according to
\begin{align}\label{eq:gov1}
 \dot{\theta}_k^{\sigma}:=\frac{\ud\theta_k^\sigma}{\ud t} &= \omega + \sum_{\tau=1}^2 \frac{K_{\sigma\tau}}{N}\sum_{l=1}^{N}\sin{[\theta_l^{\tau}-\theta_k^{\sigma}-\alpha_{\sigma\tau}]},\
\end{align}
with intrinsic frequency~$\omega$, inter- and intra-population coupling strengths~$K_{\sigma\tau}$, and phase-lag parameters $\alpha_{\sigma\tau}$, which tune between the regimes of pure sine-coupling ($\alpha_{\sigma\tau}=0$) and pure cosine-coupling ($\alpha_{\sigma\tau}=\frac{\pi}{2}$). Assuming that the populations are symmetrically coupled\cite{Abrams2008,Montbrio2004,Martens2010bistable,Martens2010var}, we define the \emph{self-} and \emph{neighbor}-coupling parameters $\alpha_s=\alpha_{11}=\alpha_{22}$, $\alpha_n=\alpha_{12}=\alpha_{21}$, and $k_s=K_{11}=K_{22}$, $k_n = K_{12}=K_{21}$. If $k_s\neq k_n$ are distinct, the system~\eqref{eq:gov1} is \emph{non-locally} coupled, an intermediate case between local (nearest-neighbor) and global (identical all-to-all) coupling. While global coupling (i.e., uniform coupling strength) can lead to chimeras in more general oscillator models~\cite{Laing2015, SethiaSen2014, Schmidt2014}, fully symmetric coupling ($K_{\sigma\tau}=K$, $\alpha_{\sigma\tau}=\alpha$) of Kuramoto-Sakaguchi phase oscillators prevents oscillators from drifting relative to each other, a feature inherent to chimera states~\cite{Ashwin1992, Ashwin2014a, Bick2015c}. Significant progress has been made in characterizing chimera states and their bifurcations in non-locally coupled populations modeled by Eqs.~\eqref{eq:gov1}~\cite{Abrams2008,Montbrio2004}, in particular with respect to their robustness towards heterogeneity in frequencies~\cite{Laing2009,Laing2012b} and network structure~\cite{Laing2012a}, and their basins of attraction~\cite{MartensPanaggioAbrams2015}. These analyses even extend to three populations~\cite{Martens2010var, Martens2010bistable}; but all are limited to networks with homogeneous phase-lags $\alpha_{\sigma\tau}=\alpha$.

Asymmetry of the phase-lag parameters $\alpha_{\sigma\tau}$ is highly relevant for real-world applications~\cite{Lohe2015}: they correspond to energy loss along transmission lines in power grids~\cite{Dorfler2009, Motter2013} and yield an approximation for periodic solutions in systems with distributed delays such as neuronal networks~\cite{Song2008} and mobile phone networks~\cite{Klinglmayr2012, Tyrrell2010}. For coupled populations of coupled phase oscillators~\eqref{eq:gov1} both the \emph{coupling strengths} and \emph{phase-lags} affect the interaction between oscillators. Mathematically speaking, since $K_{\sigma\tau}\sin{(\theta_l^{\tau}-\theta_k^{\sigma}-\alpha_{\sigma\tau})} =\Img\big(K_{\sigma\tau}e^{-i\alpha_{\sigma\tau}}e^{i(\theta_l^{\tau}-\theta_k^{\sigma})}\big)$, one can combine coupling strength and phase-lag into a single complex parameter coupling $c_{\sigma\tau}=K_{\sigma\tau}e^{-i\alpha_{\sigma\tau}}$. This yields a natural generalization of coupled populations of coupled phase oscillators~\eqref{eq:gov1} considered previously~\cite{Abrams2008,Montbrio2004,Pikovsky2008,Martens2010var,MartensThutupalli2013, MartensPanaggioAbrams2015}. In the physical context of linear coupling, as is typical of networks with mechanical or electronic coupling~\cite{MartensThutupalli2013},  one may regard this complex constant as a response function, i.e., $K_{\sigma\tau}$ and $\alpha_{\sigma\tau}$ correspond to the amplitude- and phase-response of oscillators being forced by oscillators in its own or its neighboring population. We find that coupled populations with heterogeneous~$K_{\sigma\tau}$ and~$\alpha_{\sigma\tau}$ exhibit rich dynamics, including a variety of stable uniformly synchronized, locally synchronized, as well as desynchronized states that are quite distinct from the dynamics observed for non-local coupling with identical phase-lags. 

% ==============================================
\section{Mean field description in the thermodynamic limit}

We consider the thermodynamic limit $N\rightarrow\infty$, which allows to express the ensemble dynamics in terms of the continuous oscillator density $f^{\sigma}(\theta,\omega)$. This facilitates a low-dimensional description of the dynamics via the Ott--Antonsen (OA) ansatz~\cite{Ott2008b, Ott2009} in terms of the mean-field order parameter of each population 
\[{z}^{\sigma}(t)=r_{\sigma}(t)e^{-i\phi_{\sigma}(t)}=\int e^{i\theta}f^{\sigma}(\theta,t)\ud\theta\]
with $0<r_\sigma \leq 1$.
Let $c_{s,n}=k_{s,n}e^{-i\alpha_{s,n}}$ denote the complex valued coupling parameters.
As outlined in~\Cref{app:OA}, the mean-field dynamics given by
\begin{subequations} \label{eq:z}
\begin{align}
 \label{eq:za}
 \frac{\partial \bar{z}_1}{\partial t} &=  \half \overline{(c_s{z}_1 + c_n{z}_2)} - \half (c_s z_1+c_n z_2)\bar{z}_1^2, \\
 \label{eq:zb}
 \frac{\partial \bar{z}_2}{\partial t} &=  \half  \overline{(c_s{z}_2 + c_n{z}_1)} - \half (c_s z_2+c_n z_1)\bar{z}_2^2.\
\end{align}
\end{subequations}
describe the dynamics on an invariant manifold, the OA manifold, in which the Fourier coefficients~$f_n(t)$ of the probability density~$f$ satisfy $f_{n}(t)=a(t)^n$ for some complex function $a(t)$. 
This manifold is globally attracting for a frequency distribution with non-zero width $\Delta$~\cite{Ott2009,Ott2011}. 
Studies have shown that the dynamics on the OA manifold for $n=2$ populations and sufficiently small~$\Delta$ are qualitatively the same compared to the dynamics obtained for $\Delta=0$\cite{Laing2009,Laing2012b}. 
Thus, we discuss the dynamics in the limit of  $\Delta\rightarrow 0$ using the Ott--Antonsen reduction.

\begin{figure*}[htp!]
    \centering
  \includegraphics[width=0.99\textwidth]{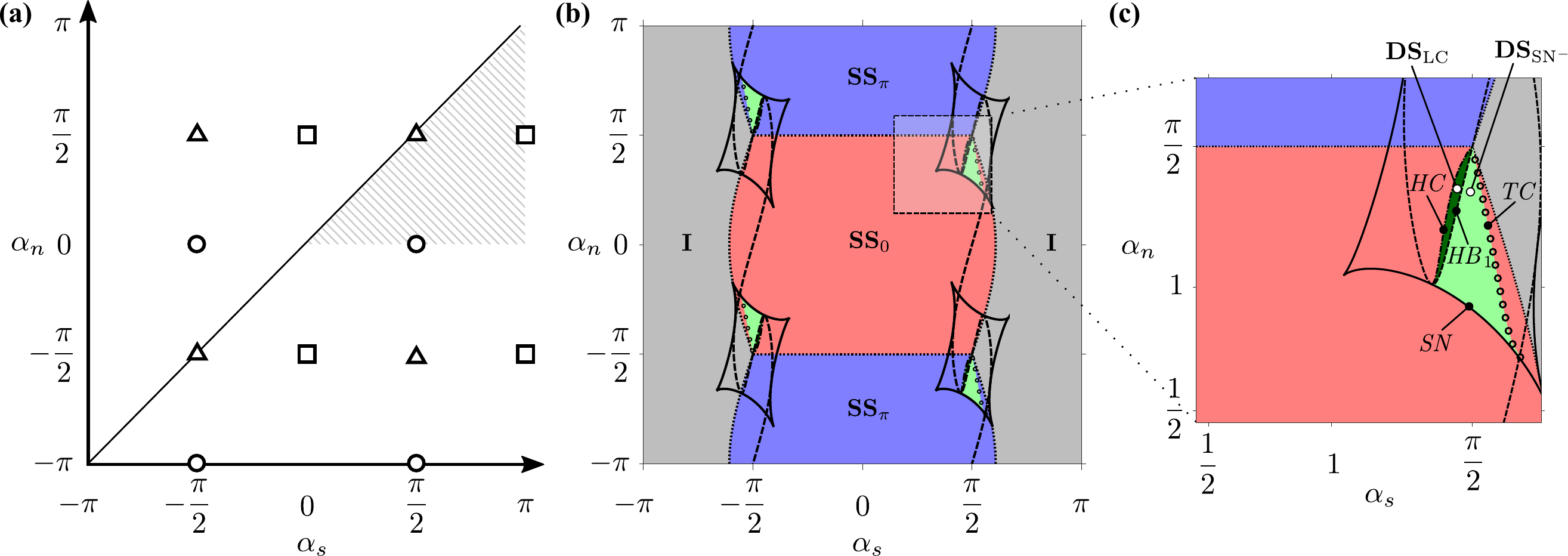}
  \caption{
  Symmetries, bifurcations, and equilibria in $(\alpha_s, \alpha_n)$-parameter space.
  {\bf (a)}: Symmetries. The diagonal indicates the parameter space for identical phase-lag parameters. Due to the parameter symmetries $\Sigma_s$, $\Sigma_n$, $\Sigma_\rho$ it suffices to consider the hatched parameter range. 
  The map $R_0$ is a time-reversing symmetry at triangles ($\vartriangle$),  and the map $R_\pi$ is a time-reversing symmetry at the circles  ($\circ$) but a regular symmetry of the system at squares ($\square$).
  {\bf (b, c)}: Bifurcation diagrams for $A=k_s-k_n=0.5$. Regions of stability are shaded in different colors: \SSzero\ (red), \SSpi\ (blue), \I\ (gray), chimera \DSone\ (green), breathing chimera \DSLC\ (dark green).  Stability boundaries 
  between the uniform states (\SSzero, \SSpi, \I) are dotted lines; transitions between non-uniform states (see right panel for close-up) are delineated by a saddle-node bifurcation ({\it SN}, solid), Hopf bifurcation (${\it HB_1}$, dashed), homoclinic bifurcation ({\it HC}, dash-dotted), and a transcritical bifurcation ({\it TC}, circles). Note that the stable regions for the chimera states \DSone\ and \DSLC\ overlap with the stable regions for either \SSzero\ or \SSpi\ and are located near the points $(\pm\frac{\pi}{2},\pm\frac{\pi}{2})$ where time-reversing symmetry $R_0$ exists.	
  Desynchronized \DD\ states with $0<r_1,r_2<0$ emerge in the transcritical bifurcation, {\it TC}.
  }
  \label{fig:figure1}
\end{figure*}

We may rewrite these equations in polar coordinates, $z_1=r_1e^{-i\phi_1}$ and $z_2=r_2e^{-i\phi_2}$.
Reducing the phase shift symmetry by introducing the phase difference $\psi=\phi_1-\phi_2$ yields the three-dimensional system
\begin{subequations}\label{eq:cylgov}
\begin{align}
 \label{eq:cylgova}
 \dot{r}_1&= \frac{1-r_1^2}{2}[k_sr_1\cos{\alpha_s}+k_nr_2\cos{(\alpha_n-\psi)}],\\
 \label{eq:cylgovb}
 \dot{r}_2&= \frac{1-r_2^2}{2}[k_sr_2\cos{\alpha_s}+k_nr_1\cos{(\alpha_n+\psi)}],\\
 \nonumber\dot{\psi}&=
 \frac{1+r_1^2}{2r_1}\left[k_sr_1\sin{\alpha_s}+k_nr_2\sin{(\alpha_n-\psi)}\right]\\
 \label{eq:cylgovc}
 &-\frac{1+r_2^2}{2r_2}\left[k_sr_2\sin{\alpha_s}+k_nr_1\sin{(\alpha_n+\psi)}\right].\
\end{align}
\end{subequations}
restricted to the cylinder \[\CC = \set{(r_1, r_2, \psi)}{0< r_1, r_2\leq 1, -\pi<\psi\leq\pi}.\]
Both complex (\ref{eq:z}) and real representation (\ref{eq:cylgov}) prove useful for the ensuing analysis.
The dynamics in Eq.~\eqref{eq:z} is conveniently displayed using the transformed variables
$\gamma = z_1 \bar{z}_2 \in \mathbb{C}, \delta = |z_1|^2-|z_2|^2 \in \mathbb{R}$, whereas~\eqref{eq:cylgov} may be represented in cylindrical coordinates (see Fig.~\ref{fig:figure5} and Ref.~\cite{MartensPanaggioAbrams2015} for examples). 

When convenient, we will rescale time and combine the coupling strength parameters~$k_n, k_s$ into a single parameter $A=k_s-k_n$, the disparity of the coupling strength between the two populations, and normalize the total coupling strength such that $k_s+k_n=1$. Note that this parametrization does not exclude the possibility of negative coupling, which corresponds to time reversal for identical frequencies, as this is equivalent to the a shift of the phase-lag parameters $(\alpha_n,\alpha_s)\mapsto(\alpha_n+\pi,\alpha_s+\pi)$, as we explain in the following.

% ==============================================
\section{Analysis}
% ==============================================

\subsection{Symmetries and Invariant Subspaces}
Symmetries imply the existence of dynamically invariant subspaces that organize the dynamics. Moreover,  parameter symmetries allow to restrict the overall parameter space. The results of this section are summarized in Fig.~\ref{fig:figure1}(a). In the following, we will write $k_s, k_n$ rather than~$A$ for ease of notation and assume the parameters to be fixed.

\paragraph{Synchronized Populations as Invariant Subspaces.}\label{sec:InvSub}
The faces of the cylinder~$\CC$ defined by 
\begin{align*}
\Ss_1 &= \set{(r_1, r_2, \psi)\in \CC}{r_1=1},\\
\Ss_2 &= \set{(r_1, r_2, \psi)\in \CC}{r_2=1},
\end{align*}
are dynamically invariant\cite{MartensPanaggioAbrams2015}. Its union $\Ss_1\cup\Ss_2$ corresponds to the points where at least one population is synchronized. The dynamics on~$\Ss_2$ (and similarly on $\Ss_1$) are given for $r=r_1$ by
\begin{subequations}\label{eq:gov2d}
\begin{align}\label{eq:gov2da}
  \dot{r}&= \dfrac{1-r^2}{2}[k_sr\cos{\alpha_s}+k_n\cos{(\alpha_n-\psi)}]\\\label{eq:gov2db}
  \dot{\psi}&=
  \dfrac{1+r^2}{2r}\left[k_sr\sin{\alpha_s}+k_n\sin{(\alpha_n-\psi)}\right]\\
  \nonumber&\qquad-k_s\sin{\alpha_s}-k_nr\sin{(\alpha_n+\psi)}.\
\end{align}
\end{subequations}
Moreover, their intersection $\SS:=\Ss_1\cap\Ss_2$ is dynamically invariant and the dynamics of~$\psi$ are given by
\begin{equation}\label{eq:gov1d}
   \dot{\psi}=k_n\left[\sin(\alpha_n-\psi)-\sin(\alpha_n+\psi)\right]
\end{equation}
and are independent of~$\alpha_s$.

\paragraph{Symmetries of the System.\label{sec:systemsymmetries}}
Recall that we have a symmetry of a dynamical system if there is a group whose action commutes with the vector field~\cite{Golubitsky2002}. Sets of points that remain fixed under the action of a subgroup of the symmetry group are dynamically invariant. 

First note that~\eqref{eq:gov1} has a continuous symmetry that acts by shifting all phases by a constant amount. Moreover, we have a permutational symmetry since oscillators within one population can be permuted as well as one can permute the populations (these two actions do not necessarily commute). As a consequence, the Ott--Antonsen equations~\eqref{eq:z} still have a phase shift symmetry as well as a symmetry that permutes the indices of the two populations. In polar coordinates with phase differences~\eqref{eq:cylgovc} the phase shift symmetry is reduced and only the permutational symmetry that acts by
\begin{equation}
\Sot :(r_1, r_2, \psi)\mapsto(r_2, r_1, -\psi)
\end{equation}
remains.
The dynamically invariant fixed point subspace is given by 
\begin{align}
\RR:=\Fix(\Sot) &= \set{(r_1, r_2, \psi)}{r_1=r_2, \psi\in\sset{0,\pi}}.\
\end{align}
which are the dynamically invariant rays described previously~\cite{MartensPanaggioAbrams2015}. On $\RR$ the dynamics for $r_1=r_2=r$ are given by
\begin{equation}\label{eq:DynInvRay}
 \dot{r} = \frac{1-r^2}{2}r(k_s\cos{\alpha_s} + k_n\cos{(\alpha_n-\psi)}).
 \end{equation}
where $\psi\in\sset{0, \pi}$. It is apparent from \eqref{eq:cylgov} that the set~$\RR$ is contained in the invariant cone $\sset{(r, r, \psi)}\subset\CC$  for $\alpha_n\in\sset{0, \pi}$ and arbitrary $\alpha_s$ and $A$.
This cone divides~$\CC$ into two dynamically invariant connected regions.
On this cone, dynamics have been studied explicitly~\cite{Martens2009, Pietras2016}.

Combining this observation with the invariant subspaces in Sec.~\ref{sec:InvSub} 
yields the existence of two equilibria, $\SSzeroI = (1, 1, 0)$ and $\SSpiI = (1, 1, \pi)$. These points are stationary since 
\[\sset{\SSzeroI, \SSpiI} = \RR \cap \SS,\]
independent of the parameters $\alpha_s, \alpha_n, k_s, k_n$. Note that for $\alpha_n\not\in\sset{\pm\frac{\pi}{2}}$ they are the only equilibria on $\SS$ since they are the only fixed points of~\eqref{eq:gov1d}.

\paragraph{Parameter and Time-reversal Symmetries.}
The system has parameter symmetries given by
\begin{align}
\label{eq:ParSymSn}
 \Sigma_n&:
 \left(\alpha_s, \alpha_n, \psi, t\right)\mapsto\left(\alpha_s, \alpha_n+\pi, \psi+\pi, t\right)\\
 \label{eq:ParSymSs}
 \Sigma_s&:
 \left(\alpha_s, \alpha_n, \psi, t\right)\mapsto\left(\alpha_s+\pi, \alpha_n, \psi+\pi, -t\right)\\
\intertext{%
where~$\Sigma_s$ also inverts time. As a consequence, we have a ``diagonal'' parameter symmetry
}
 \Sigma_{sn}&:
 \left(\alpha_s, \alpha_n, \psi, t\right)\mapsto\left(\alpha_s+\pi, \alpha_n+\pi, \psi, -t\right)
\end{align}
in $(\alpha_s, \alpha_n)$-parameter space that keeps all points of~$\CC$ fixed and inverts time.
Moreover, there is a parameter symmetry
\begin{equation}
\label{eq:ParSymSrh}
 \Sigma_\varrho:
 \left(\alpha_s, \alpha_n, \psi, t\right)\mapsto\left(\pi-\alpha_s, \pi-\alpha_n, -\psi, -t\right)
\end{equation}
which corresponds to inversion in the point $(\alpha_s, \alpha_n) = (\frac{\pi}{2}, \frac{\pi}{2})$.
To understand the dynamics on~$\CC$ it is thus sufficient to consider the set $\alpha_s, \alpha_n\in[0, \pi)$, $\alpha_s\geq\alpha_n$ -- the gray-hatched region in Fig.~\ref{fig:figure1}(a).

The parameter symmetries indicate that there are parameter values for which the system~\eqref{eq:cylgov} has time reversal symmetries. 
Since $\Sigma_\varrho$ keeps the parameter values $(\alpha_s, \alpha_n) = (\frac{\pi}{2}, \frac{\pi}{2})$ invariant, it
reduces to a time-reversal symmetry
\[R_0:(r_1, r_2, \psi)\mapsto (r_1, r_2, -\psi)\]
for these parameter values. 
Applying $\Sigma_n, \Sigma_s$, we have that $R_0$ is a time-reversal symmetry for
$(\alpha_s, \alpha_n) \in\sset{(\pm\frac{\pi}{2}, \pm\frac{\pi}{2})}$ (triangles in Fig.~\ref{fig:figure1}a)). 
This corresponds to the time-reversal symmetry in~\eqref{eq:gov1} for pure cosine coupling when the interaction between oscillators is given by an even function~\cite{Ashwin2016}. Points with $\psi\in\sset{0, \pi}$ are fixed under the action of~$R_0$. 

Similarly, parameter values that are mapped by $\Sigma_\rho$ onto their images under $\Sigma_n$ or $\Sigma_s$, give rise to symmetries or time-reversing symmetries.
The point $(\alpha_s, \alpha_n) = (\frac{\pi}{2}, 0)$ is mapped by $\Sigma_\rho$ to its image under $\Sigma_n$, that is, $\Sigma_\varrho(\frac{\pi}{2}, 0) = \Sigma_n(\frac{\pi}{2}, 0)$. This implies that we have a time-reversal symmetryok
\[R_\pi:(r_1, r_2, \psi)\mapsto (r_1, r_2, \pi-\psi)\]
for $(\alpha_s, \alpha_n) \in\sset{(\pm\frac{\pi}{2}, 0), (\pm\frac{\pi}{2}, \pi)}$ (circles in Fig.~\ref{fig:figure1}a)).
that leaves points with $\psi\in\sset{\pm\frac{\pi}{2}}$ invariant. 
Furthermore, the point $(\alpha_s, \alpha_n) = (0,\frac{\pi}{2})$ is mapped by $\Sigma_\rho$ to its image under $\Sigma_s$, that is, $\Sigma_\varrho(0,\frac{\pi}{2}) = \Sigma_s(0,\frac{\pi}{2})$. Since $\Sigma_s$ also reverses time, this implies that $R_\pi$ is a (regular) symmetry for $(\alpha_s, \alpha_n) \in\sset{(0,\pm\frac{\pi}{2}), (\pi,\pm\frac{\pi}{2})}$ (squares in Fig.~\ref{fig:figure1}a)). The invariant set $\set{(r_1,r_2,\psi)}{\psi=\pm\frac{\pi}{2}}\subset\CC$ divide phase space into two invariant regions.

% -------------------------------------------
\subsection{Full Synchrony~$\SSzeroI$, Antiphase Synchrony~$\SSpiI$, and Incoherence $\II$}

Independent of phase-lag and coupling strength, there are two equilibria where both populations are fully synchronized: in-phase synchronization $\SSzeroI = (1, 1, 0)$ and anti-phase synchronization $\SSpiI = (1, 1, \pi)$. As mentioned above, they are the only equilibria on~$\SS$ if $\alpha_n\neq\pm\frac{\pi}{2}$.

Similarly, we denote by $\II$ the equilibrium solution with $r_1=0$ and $r_2=0$ which corresponds to a completely incoherent distribution of oscillator phases~\cite{Montbrio2004} in terms of the order parameter. Note that in the finite dimensional system~\eqref{eq:gov1} the condition $r_1=0$ defines a manifold~\cite{Ashwin2016} that contains for example splay states~\cite{Swift1992,Strogatz1993} where the oscillators are evenly distributed or any other configuration that yields zero order parameter.

\paragraph{Stability of~$\SSzeroI$ and~$\SSpiI$.}
The eigenvalues of the linearization of \eqref{eq:cylgov} at \SSzero\, and \SSpi\, are
\begin{align}
 \lambda_1^{\SSzeroI}&=\lambda_2^{\SSzeroI} = - k_s\cos{\alpha_s} - k_n\cos{\alpha_n}, \label{eq:LambdaSSz1}\\
 \lambda_3^{\SSzeroI} &= -2k_n\cos{\alpha_n} \label{eq:LambdaSSz2}
\end{align}
and
\begin{align}
 \lambda_1^{\SSpiI}&= \lambda_2^{\SSpiI}= - k_s\cos{\alpha_s}+k_n\cos{\alpha_n} , \label{eq:LambdaSSp1}\\
 \lambda_3^{\SSpiI} &= 2k_n\cos{\alpha_n}  \label{eq:LambdaSSp2}
\end{align}
respectively.
The eigenvalues $\lambda_1=\lambda_2$ are degenerate, and it suffices to consider $\tau=\lambda_1+\lambda_3$ and $\Delta =\lambda_1\lambda_3$ to discuss stability. The eigenvalues are real, and thus, we can either have saddles ($\Delta<0$), unstable ($\Delta>0,\tau>0$) or stable nodes ($\Delta>0,\tau<0$). Regions of stability are  shown in \Cref{fig:stability_uniformstates}. 
Stability boundaries are located at 
\begin{align}\label{eq:stabilityboundaries}
 k_n=0, \; |\alpha_n|=\frac{\pi}{2},\; {\rm and }\; k_n=\mp k_s \dfrac{\cos{\alpha_s}}{\cos{\alpha_n}}, \
\end{align}
for \SSzero\, and \SSpi, respectively.

Since $\lambda_3^{\SSzeroI}=-\lambda_3^{\SSpiI}$, an exchange of stability occurs when $\lambda_3=0$, i.e., when $k_n=0$ or $\alpha_n=\pm\frac{\pi}{2}$.
This implies (i) $\SSzeroI$ and $\SSpiI$ always have converse stability properties for any given parameter values (unless $k_n=0$ or $\alpha_n=0$), and in particular, are never stable simultaneously and (ii) provided that the states already are stable on condition of $\lambda_1<0$,  \SSzero\, and \SSpi\, swap stability at $\lambda_3=0$.

\paragraph{Stability of~$\II$.}
Since the polar coordinates have a parameter singularity leaving~$\psi$ undefined, consider complex Eqs.~\eqref{eq:z} to determine linear stability of~\I. Separating into real and imaginary parts, we obtain the eigenvalues 
\begin{align}
  \lambda_{1,2}^{\I}&=k_{s}\cos\alpha_{s}+k_{n}\cos\alpha_{n}\pm i\left|k_{n}\sin\alpha_{n}+k_{s}\sin\alpha_{s}\right|\\
  \lambda_{3,4}^{\I}&=k_{s}\cos\alpha_{s}-k_{n}\cos\alpha_{n}\pm i\left|k_{n}\sin\alpha_{n}-k_{s}\sin\alpha_{s}\right|
\end{align}
of the Jacobian evaluated at~$\I$.
The real parts~$\Re(\lambda)$ of all four eigenvalues must be negative for this equilibrium solution to be stable. Thus we obtain the stability condition 
\[k_{s}\cos\alpha_{s}<-\left|k_{n}\cos\alpha_{n}\right|.\]  Note that $\Re(\lambda_{1,2}^{\I})=-\lambda_{1,2}^{\SSzeroI}$ and $\Re(\lambda_{3,4}^{\I})=-\lambda_{1,2}^{\SSpiI}$. It therefore follows that if either  \SSzero\, or \SSpi\, is stable, then \I\ must be unstable and vice-versa. As a consequence, in combination with the conclusions drawn previously for the fully synchronized states, we have demonstrated that $\SSzeroI$, $\SSpiI$ and~$\II$ partition parameter space into {mutually exclusively stable regions}. These regions of stability are shown in Fig.~\ref{fig:stability_uniformstates}(c).

\paragraph{Global Bifurcations and Continua of Equilibria.}
\label{sec:GlobBifEq}
The sinusoidal coupling of the system forces a degenerate bifurcation behavior that leads to mutually exclusive regions of stability. More precisely, the equilibria $\SSzeroI$, $\SSpiI$, and~$\II$ are connected by a network of invariant subspaces defined by $\SS\cup\RR$ which forces eigenvalues to always switch in pairs; see Fig.~\ref{fig:stability_uniformstates}(a).
For example, 
if $\alpha_n=\pm\frac{\pi}{2}$ then
$\lambda_3^{\SSzeroI} = \lambda_3^{\SSpiI}=0$
independently of~$A$. This implies that $\SSzeroI$ and~$\SSpiI$ swap stability in a degenerate global bifurcation with the set~$\SS$ being a continuum of equilibria as the right hand side of~\eqref{eq:gov1d} vanishes (Fig.~\ref{fig:stability_uniformstates}). 
Similarly, if $\lambda_1^{\SSzeroI} = 0$ or $\lambda_1^{\SSpiI}=0$, the right hand side of~\eqref{eq:DynInvRay} vanishes for either $\psi=0$ or $\psi=\pi$ which implies that a subset of~$\RR$ is a continuum of equilibria. Again, this yields a degenerate  bifurcation if $\SSzeroI$ or~$\SSpiI$ and~$\II$ swap stability through a continuum of equilibria. Calculating the transverse eigenvalues to the continuum of equilibria at the bifurcation (not shown) yields additional information on the dynamics close to~$\RR$ for nearby parameter values\cite{MartensPanaggioAbrams2015}.

\begin{figure}
\subfigure[$\psi=0$ plane]{\includegraphics{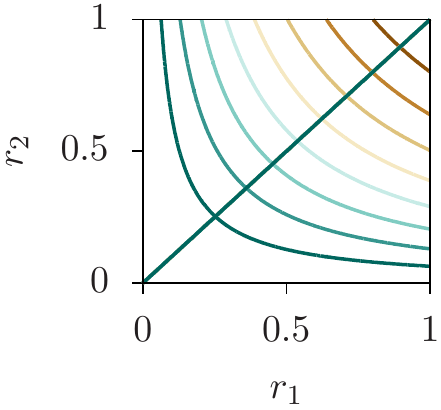}}\quad
\subfigure[$\psi=\frac{\pi}{2}$ plane]{\includegraphics{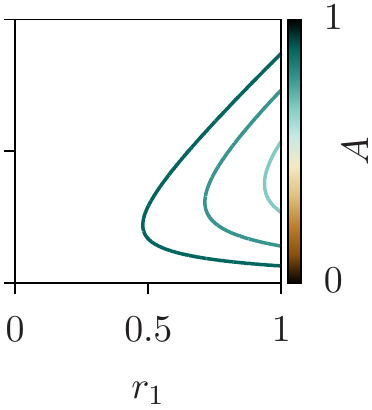}}
\caption{\label{fig:ContofEqs}Continua of equilibria for $\alpha_s=\frac{\pi}{2}$. Panel~(a) shows solutions to the algebraic equations for $\alpha_n=\frac{\pi}{2}$ and varying $A$ (line color); the diagonal is always a solution corresponding to~$\RR$. Panel~(b) depicts solutions for $\alpha_n=0$. Branches of solutions intersect the surface~$\partial \CC$ giving rise to chimera states; for small values of~$A$ solutions for $\alpha_n=0$ do not intersect~$\CC$.}
\end{figure}

\begin{figure*}
  \centering
  \includegraphics[width=0.8\textwidth]{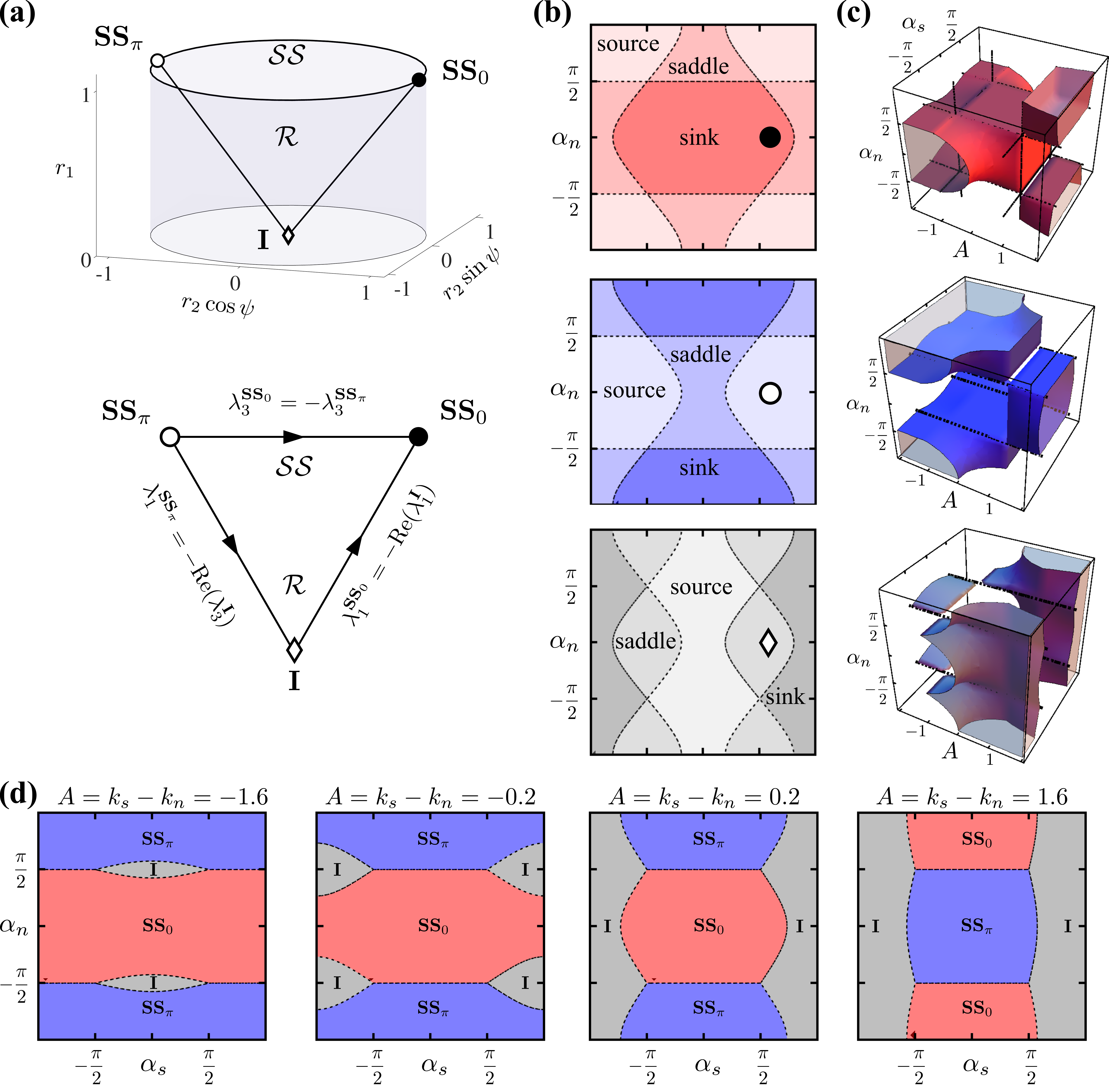}\\
  \caption{Stability of equilibria \SSzero, \SSpi, \I.
  {\bf (a):} Invariant subspaces organize the stability of the equilibria in cylinder $\CC$ (top). Network of invariant subspaces forces stability to change in global bifurcations (bottom).
  {\bf (b):} Cross-sections at $A=k_s-k_n=0.1$ divide parameter space into regions of different types of stability (dark, medium and bright shading denote regions with stable nodes, saddles and unstable nodes, respectively).
  {\bf (c):} Stable regions for the fully synchronized states $\SSzeroI$ (red), $\SSpiI$ (blue) and the fully incoherent state \I\, (gray) are shown separately.
  Dashed lines correspond to the existence of additional continua of equilibria for the time reversing symmetry $R_0$.
  {\bf (d):} The stable uniformly synchronized states, \emph{\SSzero, \SSpi\ and \I, partition parameter space into mutually exclusive regions}.
  Note that pairs of $\alpha_s=\pm\frac{\pi}{2}$, $\alpha_n=\pm\frac{\pi}{2}$ and $A=\pm 1$ are loci in parameter space where all stability regions join together: varying $A$, we always observe that the three regions join in $\alpha_s=\pm\frac{\pi}{2}$, $\alpha_n=\pm\frac{\pi}{2}$. 
  }
  \label{fig:stability_uniformstates}
\end{figure*}

Note that for parameters with time-reversal symmetries there are additional continua of equilibria in~$\CC$ that lie in the sets that remain fixed under the action of the time-reversal symmetries. More precisely, if $\alpha_s=\pm\frac{\pi}{2}$ and
\begin{equation}\label{eq:ContEq}
\alpha_n-\psi = \alpha_n+\psi = \frac{\pi}{2}\mod\pi
\end{equation} 
we have $\dot r_1 = \dot r_2 = 0$. Thus, finding equilibria of~\eqref{eq:cylgov} reduces to the algebraic condition $\dot\psi=0$ as given in~\eqref{eq:cylgovc}. If $\alpha_n = \pm\frac{\pi}{2}$ then~\eqref{eq:ContEq} is fulfilled if $\psi\in\sset{0, \pi}$. For $\alpha_s=\alpha_n = \frac{\pi}{2}$ and $\psi=0$ the condition $\dot\psi=0$ is equivalent to
\[k_s(r_2r_1^3-r_1r_2^3)+k_n(r_2^2-r_1^2)=0\]
and its solutions are depicted in Fig.~\ref{fig:ContofEqs}(a). Solutions come in pairs, that is, if $(r_1, r_2)$ is a solution so is $(r_2, r_1)$, and all points in~$\RR$ are solutions; cf.~Equation~\eqref{eq:DynInvRay}.
Similarly, if $\alpha_n\in\sset{0,\pi}$ then~\eqref{eq:ContEq} implies $\psi\in\sset{\pm\frac{\pi}{2}}$.
For $\alpha_s=\frac{\pi}{2}$, $\alpha_n=0$, and $\psi=\frac{\pi}{2}$ the condition $\dot\psi=0$ is equivalent to
\[k_s(r_2r_1^3-r_1r_2^3)-k_n(2r_1^2r_2^2+r_2^2+r_1^2)=0\]
and its solutions are depicted in Fig.~\ref{fig:ContofEqs}(b). In either case, depending on the choice of $k_s, k_n$ (or $A$) the continua of equilibria may intersect the boundary of~$\CC$ to give rise to chimera states with a neutrally stable direction.

% -------------------------------------------
\subsection{Chimera states {\texorpdfstring{\DS\,}{DS}}~and {\texorpdfstring{\SD}{SD}}}
Chimeras correspond to steady state solutions of Eq.~\eqref{eq:cylgov} on invariant surfaces $\Ss_1$ or~$\Ss_2$, where the either first population is synchronized ($r_1=1$) and the second population is partially desynchronized ($0\leq r_2<1$) or vice versa. We refer to these chimeras as \DS\ or \SD\ with subscripts to differentiate between distinct equilibria. For a given set of parameter values $(A,\alpha_s,\alpha_n)$  we find up to four branches of chimeras, three that appear to be always unstable and one that is stable in a wedge shaped region of parameter space (see Fig.~\ref{fig:figure1}). These extend the stable chimeras discussed in Ref.~\cite{Abrams2008} for identical phase-lag parameters which undergo various further bifurcations for nonidentical phase lags as discussed below.

\paragraph{Chimeras near \SSzero.}
For parameter values close to $\alpha_s=\alpha_n=\frac{\pi}{2}$, a saddle node bifurcation gives rise to two branches of equilibria on $\Ss_2$, a branch \DSone\ of equilibria that are stable close to the saddle node bifurcation and a branch \DStwo\ that is unstable close to the bifurcation. Note that these branches can change stability away from the bifurcation point as they may undergo additional bifurcations; cf.~text further below and Sec.~\ref{sec:DD}. By symmetry, analogous branches arise in $\Ss_1$.
Using perturbation theory we can approximate these states for small~$A$.  \DSone\ and \DStwo\ are described by
{\allowdisplaybreaks
\begin{align*}
\alpha_s&=\frac{\pi}{2}-A\alpha_1\\
\alpha_n&=\frac{\pi}{2}-A(\alpha_1+\Delta_1)\\
r_1&=1+A(-1\mp S)+\bigO(A^2) \\%+A^2\left(\frac{1}{2}\right)\left[-(3\Delta_1^2+11\Delta_1\alpha_1+8\alpha_1^2-2)-\frac{1}{S}(7\Delta_1^2+19\Delta_1 \alpha_1+12\alpha_1^2-2)\right]\\
r_2&=1\\
\psi&=-A(2\alpha_1+\Delta_1)+\bigO(A^2)%-A^2\alpha_1(1-S)
\end{align*}
}
where 
\[S=\sqrt{1-2\Delta_1^2-6\Delta_1\alpha_1-4\alpha_1^2}\]
and $\Delta_1$ and $\alpha_1$ are free parameters that can be independently used to set the phase-lag difference $\alpha_s-\alpha_n$ and deviation from $\frac{\pi}{2}$, respectively.  Thus, when $A$ is small, chimeras are located (approximately) along a plane in parameter space parametrized by  $\Delta_1$ and $\alpha_1$. 

\begin{figure*}[htp]
  \centering 
  \includegraphics[width=0.7\textwidth]{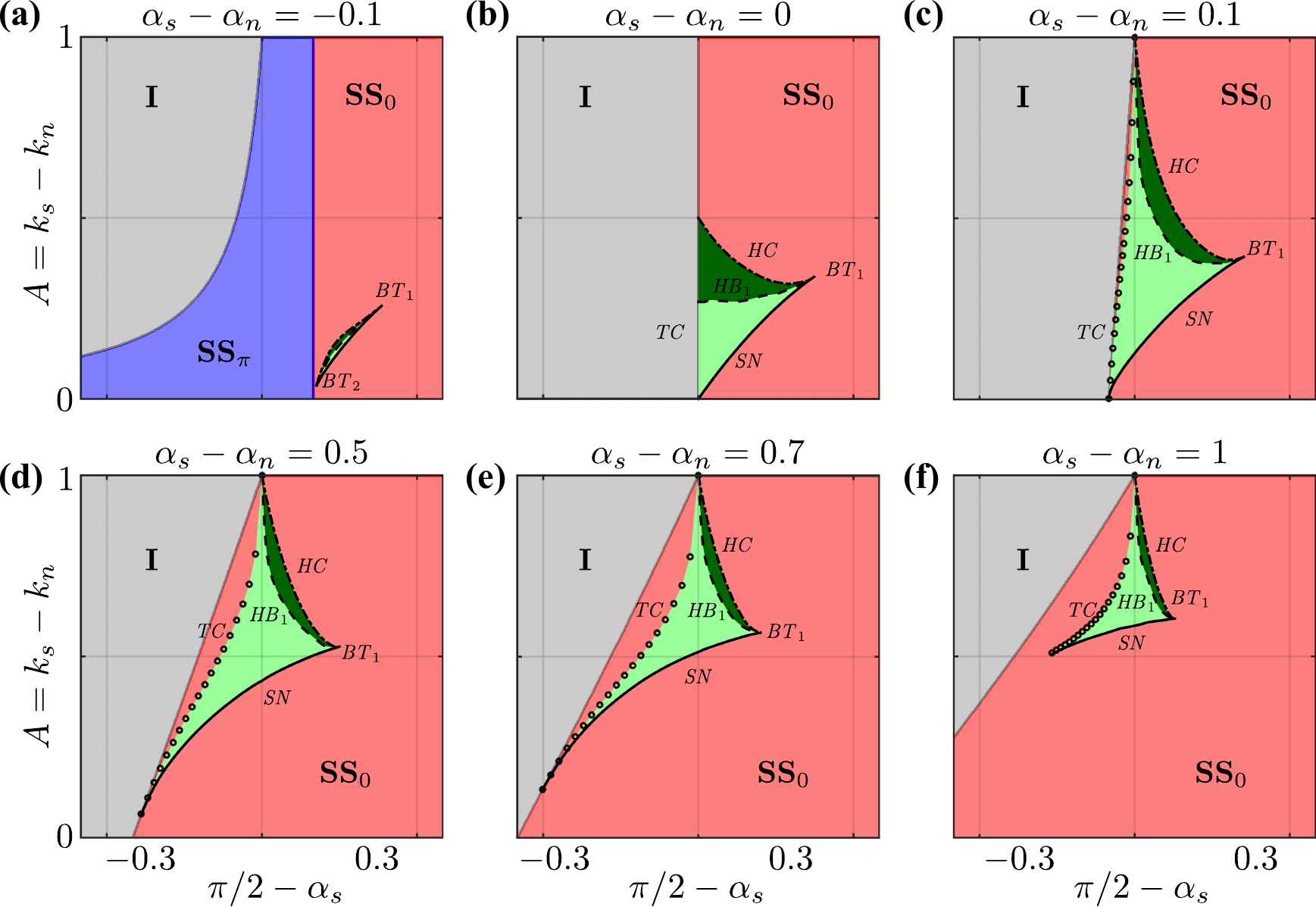}
  \caption{\label{fig:figure4}   
    Bifurcation diagrams for varying strength of phase-lag heterogeneity, $\alpha_s-\alpha_n$, allow the comparison to the case of homogeneous phase-lags~\cite{Abrams2008}. Chimera attractors reside in the light/dark green shaded regions and are bistable with \SSzero\ (or \SSpi, for the respective $\pi$-chimera). Stable chimeras (\DSone,\SDone) exist in a wedge shaped region (light green) bounded by three bifurcation curves: 
  a saddle-node bifurcation curve ({\it SN}, solid), a Hopf bifurcation curve (${\it HB_1}$, dashed), and a transcritical bifurcation curve ({\it TC}, circles). The wedge appears at $\alpha_s-\alpha_n = -0.16759$ (determined numerically) when two Bogdanov-Takens points emerge from a single point and disappears again when $\alpha_s-\alpha_n=\frac{\pi}{2}$. 
  Breathing chimeras (\DSLC, \SDLC) exist in a crescent shaped region (dark green) bounded by a Hopf bifurcation curve (${\it HB_1}$, dashed) and a homoclinic bifurcation curve ({\it HC}, dash-dotted). 
  \DD\ states emerge as the transcritical curve, {\it TC}, as one leaves the region of stable chimera states.
  Stable regions of the uniform states \SSzero, \SSpi, and \I\ are indicated by red, blue, and gray shades, respectively.
}  
\end{figure*}

Figs.~\ref{fig:figure1}(b) and \ref{fig:figure1}(c) show the regions of existence and stability for \DSone\ and \DStwo\ for fixed $A$ (see Appendix~\ref{app:bif_curves} for details).  Here we see that these chimeras exist in a bow-tie shaped region near $(\alpha_s,\alpha_n)=(\frac{\pi}{2},\frac{\pi}{2})$ bounded by the closed saddle-node curve. As one crosses this curve from the interior, \DSone\ and \DStwo\ approach each other and ultimately collide and cease to exist. Within this region \DStwo\ is always unstable, but \DSone\ is stable in a wedge shaped region with $\alpha_n<\frac{\pi}{2}$ that overlaps the region of stability of \SSzero. This stable region is bounded by curves corresponding to saddle-node, Hopf and transcritical bifurcations.   As one crosses the Hopf bifurcation, \DSone\ becomes unstable and a stable limit cycle is born that corresponds to a ``breathing chimera,'' denoted \DSLC .  This breathing chimera subsequently undergoes a homoclinic bifurcation and ceases to exist when the limit cycle collides with \DStwo. The transcritical bifurcation is discussed in Sec.~\ref{sec:DD}.

Fig.~\ref{fig:figure4} depicts these bifurcation curves for $\alpha_s-\alpha_n$ fixed. The panel with $\alpha_s-\alpha_n=0$ is equivalent to Fig.~4 in Ref.\citep{Abrams2008}.  Here we see that stable chimeras only exist for $-0.16759\leq \alpha_s-\alpha_n \leq \frac{\pi}{2}$. The saddle-node, Hopf and homoclinic bifurcation curves intersect at a Bogdanov--Takens point (${\it BT}_1$). For $\alpha_s-\alpha_n<0$, they intersect at a second Bogdanov--Takens point (${\it BT}_2$). These points merge at $\alpha_s-\alpha_n\rightarrow -0.16759$ and below this point, stable chimeras do not exist. For $\alpha_s-\alpha_n>0$, the Hopf, homoclinic and transcritical bifurcation curves intersect at the point $(A,\alpha_s)=(1,\frac{\pi}{2})$ and the transcritical and saddle-node curves also intersect, thus bounding the stability region for \DSone.  

Parameter symmetries $\Sigma_n$, $\Sigma_s$, and $\Sigma_{\varrho}$ lead to analogous chimeras in other corners of phase space.  Only \DSone\ has a stable region inside the cylinder $\CC$. Near $(\alpha_s,\alpha_n)=\pm(\frac{\pi}{2},\frac{\pi}{2})$, \DSone\ corresponds to a familiar in-phase chimera.  However, for $(\alpha_s,\alpha_n)=\pm(\frac{\pi}{2},-\frac{\pi}{2})$, \DSone\ is stable only when $\psi\approx\pi$ and therefore corresponds to an \emph{``anti-phase chimera''}.

\paragraph{Other Chimeras.}
In addition to \DSone\ and \DStwo\ we find other branches of equilibria corresponding to chimera states, denoted by \DSthree\ and \DSfour, first observed for identical phase-lag parameters~\cite{Panaggio2015b}.  The branch \DSthree\ emerges for small $A$, near \SSpi~ 
with perturbation expansion
{\allowdisplaybreaks
\begin{align*}
\alpha_{{s}}&=\frac{\pi}{2}-\sqrt {A}\alpha_1\\
\alpha_{{n}}&=\frac{\pi}{2}-\sqrt {A} \left( \alpha_1+\Delta_{{1}} \right) \\
r_1&=1+ \Delta_1\left( \Delta_{1}+\alpha_1\right) 
A+\bigO(A^2)\\
\psi&=\pi-\Delta_1\sqrt{A}+\bigO(A^{3/2})
\end{align*}
}
where $\Delta_1$ and $\alpha_1$ are again free parameters.
The branch \DSfour\ emerges when for phase-lag difference $\alpha_s-\alpha_n\approx\frac{\pi}{2}$ and phase difference $\psi\approx\frac{\pi}{2}$ between populations. More precisely,
{\allowdisplaybreaks
\begin{align*}
\alpha_{s}&=-A\alpha_1\\
\alpha_{n}&=\frac{\pi}{2}-{A}^{2}\Delta_{2}\\
r_{1}&=1 + \left( 2-\frac {{\Delta_{2}}^{2}}{2}-\alpha_1\Delta_{2} \right) {A}^{2}+\bigO(A^{2})\\
r_2&=1\\
\begin{split}
\psi&=-\frac {\pi }{2}-\left( \alpha_1+\Delta_{2} \right) A\\&\quad+\frac{\Delta_2}{2} \left(2+\frac {{\Delta_{2}}^{2}}{2} +\alpha_1\Delta_{2}\right) {A}^{2}+\bigO(A^{3}).%
\end{split}
\end{align*}
}
where $\Delta_2$ and $\alpha_1$ are analogous to the free parameters defined previously (although they occur at different orders with respect to $A$). Both \DSthree\ and \DSfour\ can be continued numerically and exist for all values of~$\alpha_s$ and~$\alpha_n$. Nonetheless, numerical evidence suggests that they are \emph{unstable} whenever they correspond to physically relevant solutions ($0\leq r_1\leq 1$), and thus we denote them with the subscript `U'.

It appears that no attracting chimera solutions exist when $k_s-k_n = A<0$ (see Fig.~\ref{fig:figure4}), i.e., neighbor coupling dominates the self-coupling, $k_n>k_s$, a rigorous proof for this observation is still missing.   

\paragraph{Absence of incoherent chimera states.} Incoherent chimera states, where one population is completely desynchronized $z_1=0$ (rather than synchronized) and $z_2\neq 0$, $\abs{z_2}<1$, only exist for specific parameter values. Suppose that $z_1=0$. Stationarity in \eqref{eq:zb} implies that $0 =  {\bar{c}_n\bar{z}_2}$, and thus we have  $z_2=0$ unless $c_n=0$.

\subsection{Desynchronized solutions \DD}\label{sec:DD}
In addition to uniformly synchronous solutions and chimeras, there are also attractors where both populations are partially synchronous or desynchronized, i.e., they satisfy $0<r_1<1$ and $0<r_2<1$ for all times. We denote such solutions by~\DD.

 \begin{figure*}[htp!]
  \centering
  \includegraphics[width=\textwidth]{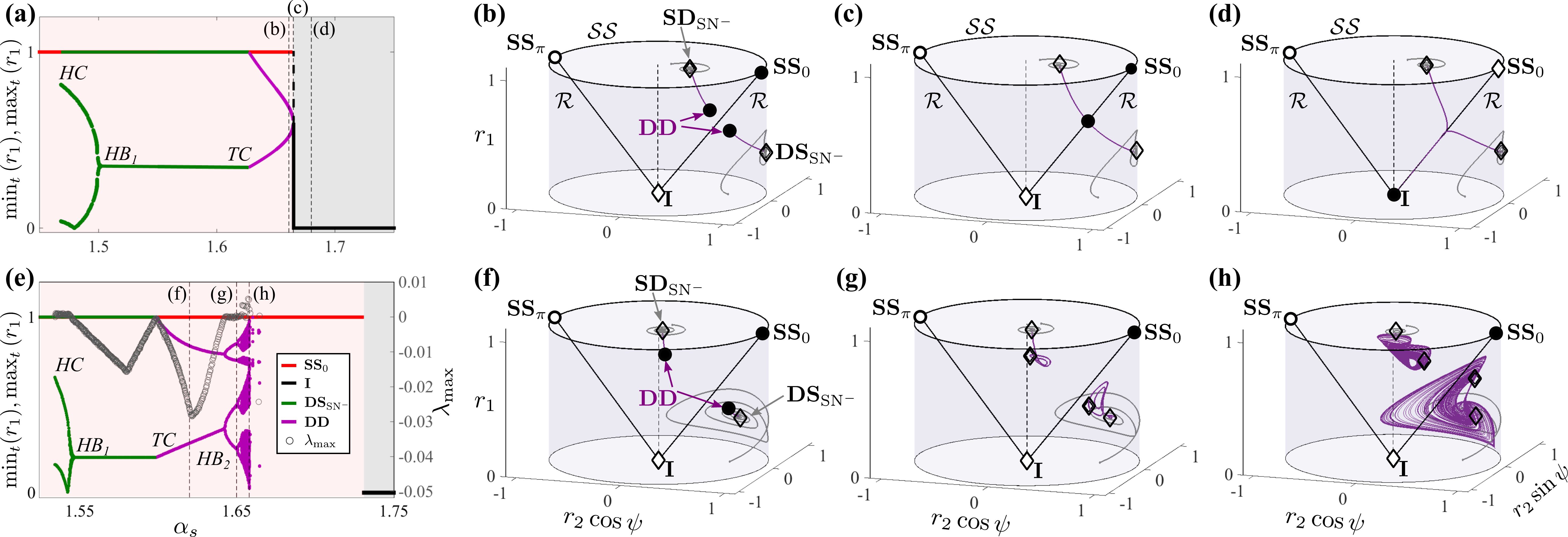}
  \caption{
  \DD\ states display a variety of bifurcation scenarios.
  {\bf (a)} The bifurcation diagram for $A=0.5$, $\alpha_n=1.2854$ reveals the following transitions, from left to right: at $\alpha_s\approx 1.47$, a breathing chimera (\DSLC) is born in a homoclinic bifurcation ($\it HC$) which becomes a stable chimera in a Hopf bifurcation ($\it HB_1$) at $\alpha_s\approx 1.5$.
  At $\alpha_s\approx 1.63$, the branch \DDeq\ penetrates the cylinder surface and swaps stability with the chimera state \SDone\ in a transcritical bifurcation ($\it TC$) -- panel {\bf (b)} ($\alpha_s=1.661$) shows a trajectory in $\CC$ initialized close to the surface of $\CC$ (gray) which passes by the chimera saddle before converging (purple) to the stable \DD\ equilibrium.
  At the global bifurcation where \SSzero\ and \I\ swap stability ($\alpha_s\approx 1.6647$), the two symmetrically related \DDeq\ branches coalesce on the corresponding continuum of equilibria on $\RR$ (panel {\bf (c)}). Panel (d) shows a trajectory converging to \I\ after the bifurcation point ($\alpha_n=1.668$).
  {\bf (e)}  Different bifurcations happen for $A=0.7$, $\alpha_n=0.44$, after the \DD\ branch gains stability (panel {\bf (f)}, $\alpha_s=1.62$) in a transcritical bifurcation ($\it TC$); the diagram shows local minima and maxima of ${r_1(t)}$ (small gray dots) and maximal Lyanpunov exponent (black large dots) after a transient transient time (see~\Cref{app:bif_curves} for details). 
  At $\alpha_s\approx 1.64$, \DDeq\ loses stability in a Hopf bifurcation ($\it HB_2$) -- {\bf (g)} shows a stable limit cycle ($\alpha_s=1.65$) and further transitions to chaos ensue~\cite{BickPanaggioMartens2016} as shown in panel {\bf (h),  $\alpha_n=1.658$}. 
  As in previous figures, fixed points in $\CC$ are shown as solid dots (stable), empty circle (unstable) and diamonds (saddles), and invariant sets~$\RR$ (line segments) and~$\SS$ (circle) are highlighted as black lines.
  Stability regions for \SSzero/\I\ are shaded in red/gray in {\bf(a,e)}.
  }
  \label{fig:figure5}
\end{figure*}

\paragraph{Equilibria.}
Branches of \DD\ equilibria can be identified by looking for stationary solutions to Eqs.~\eqref{eq:cylgov}. 
Solving for $r_1\neq \pm 1$ in \eqref{eq:cylgova} yields    
\begin{equation} \label{eq:DD_r_1}
r_1=-{\frac {k_{{n}}{r_2}\,\cos \left(\psi-\alpha_n\right) }{k_{{s}}\cos \left( \alpha_s \right) }}.
\end{equation}
Similarly, we can satisfy Eq.~\eqref{eq:cylgovb} by letting $r_2=1$ (which however, corresponds to a \DS\ chimera), or by letting   
\begin{equation} \label{eq:DD_psi}
\psi=-\frac{1}{2}\,\arccos\left( \frac{k_{s}^{2}}{k_n^2}\left[1-\cos \left( 2\,\alpha_s \right)\right] -\cos \left(2\alpha_n\right) \right)
 \end{equation}
allowing for \DD\ equilibria.
Substituting these results into \eqref{eq:cylgovc} and letting $\dot{\psi}=0$, one obtains an equation of the form 
\begin{equation}\label{eq:DD_r_2}
f(r_2,\alpha_s, \alpha_n,A)=0,
\end{equation}
which can be solved numerically.

We find that there is a branch \DDeq\ of unstable, non-physical equilibria, that is, with $0<r_1<1$ and $r_2>1$, that exists for $A = 0.7$, $\alpha_n=0.44$ and $\alpha_s\lessapprox 1.598$ (compare also with Fig.~\ref{fig:figure5}). Keeping~$A$ fixed and increasing~$\alpha_s$, \DDeq\ intersects the stable branch \DSone\ on $\Ss_1\cup\Ss_2$ in a transcritical ({\it TC}) bifurcation where the branches swap stability.
For fixed $\alpha_s-\alpha_n>0$, this curve passes through the point $(A,\alpha_s)=\left(1,\frac{\pi}{2}\right)$
with coordinates $(r_1,r_2,\psi)=(0,1,\alpha_s-\alpha_n)$ (see Figs.~\ref{fig:figure1} and \ref{fig:figure4}).
Depending on~$\alpha_n$ and~$A$, the branch \DDeq\ may disappear in a global bifurcation as it and its symmetric image collide with a continuum of equilibria. More precisely, Fig.~\ref{fig:figure5}(b-d) shows how the branch \DDeq\ collides with~$\RR$ as~$\SSzeroI$ and~$\II$ swap stability for $A=0.5$ and $\alpha_n=1.2854$ at $\alpha_s = 1.666$ (see Eq.~\eqref{eq:stabilityboundaries}). The point of intersection is given by the point on~$\RR$ where the continuum of equilibria loses transverse stability; cf.~Sec.~\ref{sec:GlobBifEq}.

\paragraph{Bifurcations to non-stationary attractors.}
Alternatively, these branches of equilibria can bifurcate to other \DD\ attractors that are contained in~$\CC$. Fig.~\ref{fig:figure5}(e) shows a numerical bifurcation diagram for varying values of~$\alpha_s$. Computing trajectories from multiple random initial conditions in~$\CC$, solutions converge to one of three types of attracting states: (i) fully synchronized solutions \SSzero\ for all values of $\alpha_s$; (ii) chimera states, \DSLC\ or \SDLC\ between ${\it HC}$ and ${\it HB_1}$ and \DSone\ or \SDone\ between ${\it HB_1}$ and ${\it TC}$; (iii) \DD\ attractors are present between ${\it TC}$ and $\alpha_s\approx 1.66$ (see \Cref{app:bif_curves} for details). 

The numerical bifurcation diagram in Fig.~\ref{fig:figure5}(e) shows that the branch of equilibria~\DDeq\ undergoes further bifurcations as~$\alpha_s$ is increased. For $\alpha_n=0.44$, $A=0.7$, \DDeq\ loses stability in a Hopf bifurcation giving a branch \DDlc\ 
of oscillatory solutions with periodic order parameters $r_1(t)$, $r_2(t)$. As $\alpha_s$ is further increased, bifurcations give rise to further complicated dynamics; details will be given in a forthcoming publication~\cite{BickPanaggioMartens2016}. Numerical calculation of maximal Lyapunov exponents~\cite{Benettin1976} indicates that some \DD\ attractors are in fact chaotic; see Fig.~\ref{fig:figure5}(h). Note that such dynamics cannot occur for \SD\ or \DS\ states as they lie on two-dimensional dynamically invariant subspaces.

% ==============================================
\section{Discussion}
% ==============================================
Heterogeneous phase-lags in populations of Kuramoto--Sakaguchi phase oscillators are crucial to understand real-world oscillatory systems. Our analysis reveals that the case of identical phase-lags is degenerate: heterogeneous phase-lags $\alpha_s\neq\alpha_n$ lead to bifurcations structures and stable equilibria not reported in systems with homogeneous phase-lags.  
For example, the shape of the triangular wedge within which stable chimeras exist (seen in Fig.~\ref{fig:figure4}) collapses when $\alpha_s<\alpha_n$. More generally, three different  cases are discernible: 
i) When $\alpha_s-\alpha_n<0$ (panel (a)) the transcritical curve is absent and instead a secondary Bogdanov-Takens point ($\it BT_2$) appears. 
ii) In contrast, when $\alpha_s-\alpha_n=0$ (panel (b)), chimeras exist in a wedge bounded by the Hopf, transcritical and homoclinic bifurcation curves, which all meet at the point $(A,\alpha_s)=(\frac{1}{2},\frac{\pi}{2})$. In this case, the transcritical bifurcation curve coincides with the boundary of the stable region for \SSzero\, and as a result, no \DD\ states are observed. 
iii) When $\alpha_s-\alpha_n>0$ (panels (c) to (f)), the intersection point for the three bifurcation curves is $(A,\alpha_s)=(1,\frac{\pi}{2})$  and the transcritical bifurcation curve is distinct from the bounding curve for the stable region for \SSzero\ leading to the existence of \DD\ equilibria with $0<r_1(t)<r_2(t)<1$. 

Furthermore, Eqs.~\eqref{eq:cylgov} with heterogeneous phase lags possess additional symmetries that allow for stable coexistence of \SSpi\ and \emph{anti-phase chimeras} (see Fig.~\ref{fig:figure1}), i.e., where the angular order parameters of the two populations are separated by approximately~$\pi$. Heterogeneous phase lags also give rise to a range of attractors where both populations are \emph{desynchronized}, $0<r_1<r_2<1$; indeed such states are absent for homogeneous phase-lag~\cite{MartensPanaggioAbrams2015}. Stable \DD\ equilibria arise through a transcritical bifurcation where they exchange stability with a chimera state on the boundary of the cylinder~$\CC$ (see also Fig.~\ref{fig:figure5}). These undergo further bifurcations yielding stable \DD\ \emph{limit cycles} (panel {\bf (g)}) and, according to our preliminary numerical investigations, \emph{chaotic attractors} (panel {\bf (h)}). In contrast to turbulence reported for continuous rings of oscillators\cite{Wolfrum2016}, the mean field equations~\eqref{eq:z} for the continuum limit of two populations of sinusoidally coupled phase oscillators~\eqref{eq:gov1} are finite-dimensional. A detailed analysis of the transition to chaos exceeds the scope of this paper and will be published elsewhere~\cite{BickPanaggioMartens2016}.

In contrast to previous studies on oscillator networks with heterogeneous phase-lag parameters, we consider heterogeneous phase-lags that preserve the permutational symmetry of the populations. Symmetry breaking heterogeneity has been considered before in a neural context where one population consists of inhibitory and the other population of excitatory elements~\cite{Maistrenko2014}. Symmetry breaking heterogeneity is similarly present in a model of two populations, one consisting of `conformists', which are experiencing positive coupling to all other oscillators, and the other one consisting of `contrarians', that experience negative coupling~\cite{Hong2011a}. The effects of symmetry breaking heterogeneity in terms of phase-lags was also studied for rings of oscillators where the phase-lag $\alpha=\alpha(x)$ is negative or positive depending on the position on the ring~\cite{Zhu2013}.

\paragraph{Persistence of chimera states.}
In contrast to a discrete ring of finitely many oscillators, chimera states in systems of finite populations of oscillators appear to be a persistent (rather than transient) phenomenon. Chimera states on discretizations of rings of oscillators with sinusoidal coupling between oscillators have been reported to have a finite lifetime that increases like a power law with system size~\cite{Wolfrum2011b}. 
Recent numerical simulations have indicated that this lifetime can be extended by considering generalized coupling where the coupling function has higher nontrivial harmonics~\cite{Suda2015}, similar to weak chimeras in small networks of oscillators~\cite{Ashwin2014a, Panaggio2016, Bick2015c} where one can prove the existence of asymptotically stable dynamically invariant sets. By contrast, extensive computational analysis of chimera states in the finite-size system Eqs.~\eqref{eq:gov1} with identical phase-lags and sinusoidal coupling displays no transient behavior~\cite{Olmi2015,OlmiMartens2015}\footnote{Refs.~\cite{Olmi2015,OlmiMartens2015} mainly concern the `Kuramoto model with inertia', but the case of zero inertia which amounts to Eqs.~\eqref{eq:gov1} is also treated.}. These simulations were limited to the case that at least one population is synchronized, and the question whether this is also true for the variety of \DD\ solutions remains to be explored.

\paragraph{Symmetries.}
Eqs.~\eqref{eq:gov1} obey various symmetries that we have investigated in detail. 
Two symmetries, $\Sigma_s, \Sigma_n$, helped in particular to simplify our analysis, as they imply that we may restrict our attention to the parameter region $0<\alpha_s,\alpha_n<\pi$. Applying these symmetry operations to chimeras and the \DD\ state explains how analogous states emerge in four distinct corners of parameter space, namely $|\alpha_s|=\frac{\pi}{2}, |\alpha_n|=\frac{\pi}{2}$, as shown in Fig.~\ref{fig:figure1}(b).
We have bistability between stable chimeras with $\psi\approx 0$ and the stable equilibrium \SSzero\ near $(\alpha_s,\alpha_n)=\pm (\frac{\pi}{2},\frac{\pi}{2})$ and, similarly, bistability between stable chimeras with $\psi\approx\pi$ and \SSpi\ near $(\alpha_s,\alpha_n)=\pm (\frac{\pi}{2},-\frac{\pi}{2})$. These anti-phase chimeras are unstable with homogeneous phase-lags, but they have been observed in experiments involving coupled metronomes where they also coexist with a uniform anti-phase state~\cite{MartensThutupalli2013}. 
In other words, stable chimeras only exist near the points in parameter space where the uniformly synchronized states, \SSzero, \SSpi, and \I\, are all neutrally stable.  This suggests that these partially synchronized dynamics represent a state of compromise between `nearly' stable equilibria, thus supporting the picture of chimera states emerging in a competition of fully synchronized states~\cite{MartensThutupalli2013}.  This compromise is reminiscent of stable or moving fronts between bistable equilibria in nonlinear PDEs \cite{Cross1993}.  
The significance of the four parameter combinations $(\alpha_s,\alpha_s)=(\pm \frac{\pi}{2},\pm \frac{\pi}{2})$ is made evident further due to the presence of continua of equilibria, which -- depending on the particular coupling strength -- may intersect the boundary $r_1=1$ (or $r_2$) and give rise to chimera states. 

\paragraph{Resonance.}
We have mentioned that it is possible to interpret the coupling strength and phase-lag, $K_{\sigma\tau}$ and $\alpha_{\sigma\tau}$, as amplitude- and phase-responses in a forced oscillator system. In this context, inter- and intra-population coupling terms provide `forcing'. It is interesting to remark that the parameter values $\alpha_s = \frac{\pi}{2}$ and $\alpha_n= \frac{\pi}{2}$ are reminiscent of resonance points. When $\alpha_s\approx \frac{\pi}{2}$, oscillators stand in resonance with other oscillators from the same population -- similarly, oscillators resonate with the neighboring oscillator population when $\alpha_n \approx  \frac{\pi}{2}$. 
The hypothesis that resonance may play an important role in generating chimera states was first mentioned in an experimental and theoretical study on chimera states emerging in a system of coupled mechanical oscillators~\cite{MartensThutupalli2013} (see also~\cite{Panaggio2016}). In the experiment, metronomes served as mechanical limit-cycle oscillators and the mechanical coupling was mediated through a mass-spring-friction system. It was observed that chimera states and partly desynchronized states (\DD) occur when oscillators and the coupling medium are near resonance. Notably, the parameter region where these states arise is includes the boundary between regions with uniformly synchronized states \SSzero\ and \SSpi\ -- in agreement with predictions of the Newtonian model describing this system~\cite{MartensThutupalli2013}. 
Further analysis and a more detailed exploration of the relationship between this experiment (and its Newtonian model) and the model presented here will be discussed in a forthcoming paper.

\paragraph{Outlook and perspectives.}
We anticipate that further understanding of the dynamics of coupled phase oscillators will shed light on the synchronization properties of real-world oscillatory systems and exciting questions remain.  For instance, what is the size and the shape of the basins of attraction of chimera states and desynchronized states in the presence of heterogeneous phase-lags and how do they deform as the parameters are varied? Moreover, % Setting the stage for $n$ populations:
while we only considered the dynamics of two populations of phase oscillators with heterogeneous phase-lags, our results suggest that the dynamics are equally rich for system consisting of more than two populations. Previous studies~\cite{Martens2010bistable,Martens2010var,Shanahan2010,Wildie2012} have only considered homogeneous phase-lags. How our results generalize to multiple populations and what novel dynamics are possible is a question that will be addressed in future research.

% ==============================================

\begin{acknowledgments}
We thank Shashi Thutupalli for useful discussions, and the editors of this focus issue, Danny Abrams, Lou Pecora and Adilson Motter for organizing this timely publication. 
Research conducted by EAM is supported by the Dynamical Systems Interdisciplinary Network, University of Copenhagen. CB has received funding from the People Programme (Marie Curie Actions) of the European Union’s Seventh Framework Programme (FP7/2007–2013) under REA grant agreement no.~626111.
\end{acknowledgments}

% ==============================================
\appendix
\section{Ott--Antonsen reduction}\label{app:OA}
\paragraph{Derivation.}
Let us consider the Kuramoto-Sakaguchi model with non-local coupling between two populations~\cite{Abrams2008,Montbrio2004,MartensPanaggioAbrams2015} of~$N$ oscillators,
\begin{align}\label{eq:gov11}
 \dot{\theta_k^{\sigma}} &= 
 \omega_k^\sigma + \sum_{\tau=1}^2 \frac{K_{\sigma\tau}}{N}\sum_{l=1}^{N}
 \sin{(\theta_l^{\tau}-\theta_k^\sigma-\alpha_{\sigma\tau})},\
\end{align}
where $\theta_k^{\sigma}$ is the phase of the $k$th oscillator, $k=1, \ldots, N$, of population $\sigma = 1,2$.

To study the mean field dynamics, we consider the thermodynamic limit where $N\rightarrow \infty$. This allows for a description of the dynamics in terms of the mean-field order parameter~\cite{Ott2008b,Ott2009,Ott2011}. 
We define two order parameters for each population $\sigma=1,2$,
\begin{align}
 z_\sigma(t)&= \int_{-\infty}^{\infty}\int_0^{2\pi}e^{i\theta^\sigma}f_\sigma(\omega^\sigma,\theta^\sigma,t)\ud\theta^\sigma \ud\omega^\sigma,\
\end{align}
where $f_\sigma(\omega^\sigma,\theta^\sigma,t)$ is the probability density of oscillators in population $\sigma$, obeying 
the continuity equation
\begin{align}
 \frac{\partial f_{\sigma}}{\partial t} + \frac{\partial}{\partial\theta}(f_\sigma v_\sigma)=0,\
\end{align} 
where $v_\sigma(\omega^\sigma,\theta^\sigma,t)$ is their velocity, given by
\begin{align}
 \nonumber v_\sigma &= 
 \omega^\sigma
 +
 \sum_{\tau=1}^2
 K_{\sigma\tau}
 \int_{-\infty}^{\infty}
 \int_0^{2\pi}
 f_\tau(\omega^\tau,\theta^\tau,t)
 \\
 &\qquad\times \sin(\theta^{\tau}-\theta^{\sigma}-\alpha_{\sigma\tau})
 \ud\theta^{\tau}\ud\omega^{\tau}\\
&=  \omega^\sigma  +\sum_{\tau=1}^2\frac{K_{\sigma\tau}}{2i}[z_{\tau}e^{-i(\theta^{\sigma}+\alpha_{\sigma\tau})}-\bar{z}_{\tau}e^{i(\theta^{\sigma}+\alpha_{\sigma\tau})}
].\
\end{align}

Following Ott and Antonsen~\cite{Ott2008b,Ott2009}, we consider probability densities along a manifold given by 
\begin{align}
  f_\sigma=\frac{g_\sigma(\omega^\sigma)}{2\pi}\left[1+\sum_{n=1}^\infty \left(a_{\sigma}(\omega^\sigma,t)e^{i\theta}\right)^n +c.c.\right].
\end{align}
Using this ansatz, we find the dynamics governed by a partial (integro-)differential equations of the form
\begin{align}\label{eqn_a}
0&=  \frac{\partial a_\sigma}{\partial t} + i\omega^\sigma a_\sigma - \half\sum_{\tau=1}^2 K_{\sigma\tau}\left[
e^{i\alpha_{\sigma\tau}}\bar{z}_{\tau} 
-e^{-i\alpha_{\sigma\tau}}  z_{\tau}a_\sigma^2
\right]\
\end{align}
where
\begin{align}\label{eq:zint}
 z_\sigma(t)&=\int_{-\infty}^{\infty} \bar{a}_\sigma(\omega^\sigma,t)g_\sigma(\omega^{\sigma})\ud\omega^\sigma.\
\end{align}
The latter integral solves by choosing a Lorentzian distribution 
\begin{align}
 g_\sigma(\omega^\sigma)&=\frac{\Delta_\sigma/\pi}{(\omega^{\sigma}-\Omega_\sigma)+\Delta_\sigma^2},\
\end{align}
with centers $\Omega_\sigma$ and width (half width at half maximum)
 $\Delta_\sigma$. Then, we have $z_\sigma(t)=\bar{a}_\sigma(\Omega_\sigma-i\Delta_\sigma,t)$ and evaluating \eqref{eqn_a} and \eqref{eq:zint} at the poles $\omega^{\sigma}=\Omega_\sigma-i\Delta_\sigma$, we obtain
\begin{align}
 \nonumber\frac{\partial \bar{z}_\sigma}{\partial t} &= - (\Delta_\sigma+i\Omega_\sigma)\bar{z}_\sigma\\
 &\qquad+\half\sum_{\tau=1}^2 K_{\sigma\tau}\left[e^{i\alpha_{\sigma\tau}}\bar{z}_{\tau} -e^{-i\alpha_{\sigma\tau}} z_{\tau}\bar{z}_\sigma^2\right],\\
  =&- (\Delta_\sigma+i\Omega_\sigma)\bar{z}_\sigma +\half\sum_{\tau=1}^2 \left[\bar{c}_{\sigma\tau}\bar{z}_{\tau} - c_{\sigma\tau} z_{\tau}\bar{z}_\sigma^2\right].\
\end{align}

\paragraph{The limit of identical frequencies.}

The Ott--Antonsen (OA) manifold, in which the Fourier coefficients $f_n(t)$ of the probability density $f$ satisfy $f_{n}(t)=a(t)^n$, is globally attracting for a frequency distribution with non-zero width $\Delta$~\cite{Ott2009,Ott2011}. For identical oscillators ($\Delta_\sigma=0,\Omega_1=\Omega_2$), the dynamics for the problem can be described by reduced equations using the Watanabe-Strogatz ansatz~\cite{WatanabeStrogatz1994}, as shown in Pikovsky and Rosenblum~\cite{Pikovsky2008}; the authors showed that Eqs.~\eqref{eq:gov11} may also be subject to more complicated dynamics than those described by the OA ansatz. Studies by Laing~\cite{Laing2009,Laing2012b} investigated the dynamics using the OA ansatz for $n=2$ populations for the case of non-identical frequencies and found that the dynamics for sufficiently small $\Delta$ is qualitatively equivalent to the dynamics obtained for $\Delta=0$. It is therefore justified to discuss the dynamics for $\Delta\rightarrow 0$ representing the case of \emph{nearly} identical oscillators using the OA reduction.
The limit of identical frequencies means that we let $g_\sigma(\omega)\mapsto\delta(\omega^\sigma-\Omega)$, implying that 
$z_\sigma(t)=\bar{a}_\sigma(t)$, and the governing equations reduce to 
\begin{align}\label{eq:zgov0}
 \frac{\partial \bar{z}_\sigma}{\partial t} 
 &=
 \half
 \sum_{\tau=1}^2 \left(\bar{c}_{\sigma\tau}\bar{z}_{\tau} - {c_{\sigma\tau}} z_{\tau}\bar{z}_\sigma^2\right),\
\end{align}
which are equations~\eqref{eq:za} and \eqref{eq:zb}.

\section{Bifurcation Curves and Stability}\label{app:bif_curves}
The bifurcation curves in Fig.~\ref{fig:figure1} were obtained following an approach similar to the one outlined in Ref.~\citep{Abrams2008}. We first compute the Jacobian $J$ of Eq.~\eqref{eq:cylgov} at a fixed point.  Fixed points \DSone\ and \DStwo\ satisfy Eq.~\eqref{eq:DD_r_1},  $r_2=1$, and $\dot{\psi}=0$ in Eq.~\eqref{eq:cylgovc}.  The saddle-node bifurcation curve where these two chimeras coincide can be obtained by solving $\det{(J)}=0$.  The Hopf bifurcation can be computed in a similar manner by setting ${\rm tr}{(J)}=0$ with the same fixed points as above.   The transcritical bifurcation curve satisfies \eqref{eq:DD_r_2} with $r_2=1$,  and $r_1$ and $\psi$ given by Eqs.~\eqref{eq:DD_r_1} and \eqref{eq:DD_psi} respectively. These bifurcation curves were computed via numerical continuation in MATCONT and verified by inspection of phase portraits corresponding to Eq.~\eqref{eq:cylgov}. The stability of chimeras was confirmed by numerically computing the eigenvalues of the Jacobian and by numerically integrating Eq.~\eqref{eq:cylgov}. 

The bifurcation curves {\it SN}, {\it HB}, {\it HC} and {\it TC} in Fig.~\ref{fig:figure4} were determined by inspection of phase portraits by considering Eqs.~\eqref{eq:cylgov} on the invariant surface defined by $r_2 = 1$ while observing eigenvalues of the full three dimensional system defined by Eqs.~\eqref{eq:cylgov}. Bogdanov Takens points (${\it BT_1,BT_2}$) were numerically computed by solving fixed point conditions of Eqs.~\eqref{eq:cylgov} simultaneously with the conditions for saddle-node ($\det{(J)}=0$) and Hopf bifurcations (${\rm tr}{(J)}=0$ and $\det{J}>0$),  where $J$ denotes the Jacobian of \eqref{eq:cylgov}. Similarly, the intersection point of the {\it SN } and {\it TC} curves for $\alpha_s>\alpha_n$ were numerically determined by solving the fixed point, saddle-node and transcritical conditions; the latter is determined by observing when one of the eigenvalues of \SDone\ (\DSone) is zero.
The intersection point at $(A,\alpha_s)=(1,\frac{\pi}{2})$ for $\alpha_s>\alpha_n$ was determined by simultaneously solving the {\it SN } and {\it TC} conditions.

Fig.~\ref{fig:figure5} (d) is computed by numerically continuing the branches \SSzero, \SD, \DD\ and \I. 
Fig.~\ref{fig:figure5} (h) samples trajectories of \eqref{eq:cylgov} for a given number of random initial conditions (uniformly drawn from $0<r_{1,2}<1, -\pi<\psi<\pi$), which converge to either of the following three attracting states: \SSzero, \SD\ (or \DS) or \DD. After a transient time of $T=4000$,  we report temporal local minima and maxima of ${r_1(t)}$ in time, measured over a time period of $T'=2000$.

\end{document}